\def\be{\begin{equation}}
\def\ee{\end{equation}}
\def\ba{\begin{array}}
\def\ea{\end{array}}

\def\dps{\displaystyle}

\documentclass[aps,amsmath,amssymb,amsfonts,showpacs,twocolumn,pra]{revtex4}
\usepackage{graphicx}
\usepackage{epstopdf}
\def\qed{\leavevmode\unskip\penalty9999 \hbox{}\nobreak\hfill
     \quad\hbox{\leavevmode  \hbox to.77778em{%
               \hfil\vrule   \vbox to.675em%
               {\hrule width.6em\vfil\hrule}\vrule\hfil}}
     \par\vskip3pt}

\begin{document}
\title{Tighter entanglement monogamy relations of qubit systems}
\author{Zhi-Xiang Jin$^{1}$}
\author{Shao-Ming Fei$^{1,2}$}

\affiliation{$^1$School of Mathematical Sciences, Capital Normal University,
Beijing 100048, China\\
$^2$Max-Planck-Institute for Mathematics in the Sciences, 04103 Leipzig, Germany}

\begin{abstract}

Monogamy relations characterize the distributions of entanglement in multipartite systems.
We investigate monogamy relations related to the concurrence $C$ and the entanglement of formation $E$.
We present new entanglement monogamy relations satisfied by the $\alpha$-th power of concurrence 
for all $\alpha\geq2$, and the $\alpha$-th power of the entanglement of formation for all $\alpha\geq\sqrt{2}$. These monogamy relations are shown to be tighter than the existing ones.
\end{abstract}

\maketitle

\section{Introduction}
Quantum entanglement \cite{MANC,HPMK,FMA,CAF,HPBS,HPBO,JIDV,ZZZG} is an essential feature of quantum mechanics. As one of the fundamental differences between quantum entanglement and classical correlations, a key property of entanglement is that a quantum system entangled with one of other subsystems limits its entanglement with the remaining ones. The monogamy relations give rise to the distribution of entanglement in the multipartite setting. Monogamy is also an essential feature allowing for security in quantum key distribution \cite{MP}.	
  									
For a tripartite system $A$, $B$ and $C$, the usual monogamy of an entanglement measure $\mathcal{E}$ implies that \cite{MK} the entanglement between $A$ and $BC$ satisfies $\mathcal{E}_{A|BC}\geq \mathcal{E}_{AB} +\mathcal{E}_{AC}$. Such monogamy relations are not always satisfied by all entanglement measures for all quantum states. It has been shown that the squared concurrence $C^2$ \cite{TJ,YKM} and the squared entanglement of formation $E^2$ \cite{TR} satisfy the monogamy relations for multi-qubit states.
It is further proved that \cite{ZXN} $C^{\alpha}$ and $E^{\alpha}$ satisfy the monogamy inequalities for $\alpha\geq2$ and $\alpha\geq\sqrt{2}$, respectively.

In this paper, we show that the monogamy inequalities obtained so far can be made tighter. We establish entanglement monogamy relations for the $\alpha$-th power of the concurrence $C$ and the entanglement of formation $E$ which are tighter than those in \cite{ZXN}, which give rise to finer characterizations of the
entanglement distributions among the multipartite qubit states.

\section{TIGHTER MONOGAMY RELATION OF CONCURRENCE}

We first consider the monogamy inequalities related to concurrence. Let $H_X$ denote a discrete finite dimensional complex vector space associated with a quantum subsystem $X$.
For a bipartite pure state $|\psi\rangle_{AB}$ in vector space $H_A\otimes H_B$, the concurrence is given by \cite{AU,PR,SA}
\begin{equation}\label{CD}
C(|\psi\rangle_{AB})=\sqrt{{2\left[1-\mathrm{Tr}(\rho_A^2)\right]}},
\end{equation}
where $\rho_A$ is the reduced density matrix by tracing over the subsystem $B$, $\rho_A=\mathrm{Tr}_B(|\psi\rangle_{AB}\langle\psi|)$. The concurrence for a bipartite mixed state $\rho_{AB}$ is defined by the convex roof extension
\begin{equation*}
 C(\rho_{AB})=\min_{\{p_i,|\psi_i\rangle\}}\sum_ip_iC(|\psi_i\rangle),
\end{equation*}
where the minimum is taken over all possible decompositions of $\rho_{AB}=\sum_ip_i|\psi_i\rangle\langle\psi_i|$, with $p_i\geq0$ and $\sum_ip_i=1$ and $|\psi_i\rangle\in H_A\otimes H_B$.

For an $N$-qubit pure state $|\psi\rangle_{AB_1\cdots B_{N-1}}\in H_A\otimes H_{B_1}\otimes\cdots\otimes H_{B_{N-1}}$, the concurrence $C(|\psi\rangle_{A|B_1\cdots B_{N-1}})$ of the state $|\psi\rangle_{A|B_1\cdots B_{N-1}}$, viewed as a bipartite state under the partitions $A$ and $B_1,B_2,\cdots, B_{N-1}$, satisfies the Coffman-Kundu-Wootters (CKW) inequality \cite{TJ,YKM},
\begin{equation}\label{C2}
  C^2_{A|B_1,B_2\cdots,B_{N-1}}\geq C^2_{A|B_1}+C^2_{A|B_2}+\cdots+C^2_{A|B_{N-1}},
\end{equation}
where $C_{AB_i}=C(\rho_{AB_i})$ is the concurrence of $\rho_{AB_i}=\mathrm{Tr}_{B_1\cdots B_{i-1}B_{i+1}\cdots B_{N-1}}(|\psi\rangle_{AB_1\cdots B_{N-1}}\langle\psi|)$, $C_{A|B_1,B_2\cdots,B_{N-1}}=C(|\psi\rangle_{A|B_1\cdots B_{N-1}})$. It is further proved that for $\alpha\geq2$, one has \cite{ZXN},
\begin{equation}\label{CA}
 C^{\alpha}_{A|B_1,B_2\cdots,B_{N-1}}\geq C^{\alpha}_{A|B_1}+C^{\alpha}_{A|B_2}+\cdots+C^{\alpha}_{A|B_{N-1}}.
\end{equation}

In fact, as the characterization of the entanglement distribution among the subsystems, the monogamy inequalities satisfied by the concurrence can be refined and becomes tighter. Before finding tighter monogamy relations of concurrence, we first introduce a Lemma.

{[\bf Lemma]}. For any $2\otimes2\otimes2^{n-2}$ mixed state $\rho\in H_A\otimes H_{B}\otimes H_{C}$, if $C_{AB}\geq C_{AC}$, we have
\begin{equation}\label{CK1}
  C^\alpha_{A|BC}\geq  C^\alpha_{AB}+\frac{\alpha}{2}C^\alpha_{AC},
\end{equation}
for all $\alpha\geq2$.

{[\sf Proof]}. For arbitrary $2\otimes2\otimes2^{n-2}$ tripartite state $\rho_{ABC}$, one has \cite{TJ,XJ}, $C^2_{A|BC}\geq C^2_{AB}+C^2_{AC}.$
If $C_{AB}\geq C_{AC}$, we have
\begin{eqnarray*}
  C^\alpha_{A|BC}&&\geq (C^2_{AB}+C^2_{AC})^{\frac{\alpha}{2}}=C^\alpha_{AB}\left(1+\frac{C^2_{AC}}{C^2_{AB}}\right)^{\frac{\alpha}{2}} \\
   && \geq C^\alpha_{AB}\left[1+\frac{\alpha}{2}\left(\frac{C^2_{AC}}{C^2_{AB}}\right)^{\frac{\alpha}{2}}\right]=C^\alpha_{AB}+\frac{\alpha}{2}C^\alpha_{AC},
\end{eqnarray*}
where the second inequality is due to the inequality $(1+t)^x\geq 1+xt \geq 1+xt^x$ for $x\geq1,~0\leq t\leq1$. \qed

In the Lemma, without loss of generality, we have assumed that $C_{AB}\geq C_{AC}$, since the subsystems
$A$ and $B$ are equivalent. Moreover, in the proof of the Lemma we have assumed $C_{AB}>0$.
If $C_{AB}=0$ and $C_{AB}\geq C_{AC}$, then $C_{AB}=C_{AC}=0$. The lower bound is trivially zero.
For multipartite qubit systems, we have the following Theorem.

{[\bf Theorem 1]}. For any $2\otimes2\otimes\cdots\otimes2$ mixed state $\rho\in H_A\otimes H_{B_1}\otimes\cdots\otimes H_{{B_{N-1}}}$, if
${C_{AB_i}}\geq {C_{A|B_{i+1}\cdots B_{N-1}}}$ for $i=1, 2, \cdots, m$, and
${C_{AB_j}}\leq {C_{A|B_{j+1}\cdots B_{N-1}}}$ for $j=m+1,\cdots,N-2$,
$\forall$ $1\leq m\leq N-3$, $N\geq 4$, we have
\be\label{MC}
\begin{array}{l}
\dps C^\alpha_{A|B_1B_2\cdots B_{N-1}}\geq C^\alpha_{A|B_1}\\[3mm]
~~~~~\dps +\frac{\alpha}{2} C^\alpha_{A|B_2}+\cdots+\left(\frac{\alpha}{2}\right)^{m-1}C^\alpha_{A|B_m}\\[3mm]
~~~~~\dps +\left(\frac{\alpha}{2}\right)^{m+1}(C^\alpha_{A|B_{m+1}}
 +\cdots+C^\alpha_{A|B_{N-2}})\\[3mm]
~~~~~\dps +\left(\frac{\alpha}{2}\right)^{m}C^\alpha_{A|B_{N-1}}
\end{array}
\ee
for all $\alpha\geq2$.

{\sf [Proof].} By using the inequality (\ref{CK1}) repeatedly, one gets
\be\label{CAK1}
\ba{l}
 \dps C^{\alpha}_{A|B_1B_2\cdots B_{N-1}}\geq  C^{\alpha}_{A|B_1}+\frac{\alpha}{2}C^{\alpha}_{A|B_2\cdots B_{N-1}}\\[3mm]
 \dps~~~~ \geq C^{\alpha}_{A|B_1}+\frac{\alpha}{2}C^{\alpha}_{A|B_2}
 +\left(\frac{\alpha}{2}\right)^2C^{\alpha}_{A|B_3\cdots B_{N-1}}\\[3mm]
 \dps~~~~ \geq\cdots\geq C^{\alpha}_{A|B_1}+\frac{\alpha}{2}C^{\alpha}_{A|B_2}
  +\cdots+\left(\frac{\alpha}{2}\right)^{m-1}C^{\alpha}_{A|B_m}\\[3mm]
 \dps~~~~~~~~~~~~~~~~ +\left(\frac{\alpha}{2}\right)^m C^{\alpha}_{A|B_{m+1}\cdots B_{N-1}}.
\ea
\ee
As ${C_{AB_j}}\leq {C_{A|B_{j+1}\cdots B_{N-1}}}$ for $j=m+1,\cdots,N-2$, by (\ref{CK1}) we get
\begin{eqnarray}\label{CAK2}
C^{\alpha}_{A|B_{m+1}\cdots B_{N-1}}\geq \frac{\alpha}{2}C^{\alpha}_{A|B_{m+1}}+C^{\alpha}_{A|B_{m+2}\cdots B_{N-1}}\nonumber\\
\geq \frac{\alpha}{2}(C^{\alpha}_{A|B_{m+1}}+\cdots+C^{\alpha}_{A|B_{N-2})}+C^{\alpha}_{A|B_{N-1}}.
\end{eqnarray}
Combining (\ref{CAK1}) and (\ref{CAK2}), we have Theorem 1. \qed

As for $\alpha\geq 2$, $(\alpha/2)^m\geq 1$ for all $1\leq m\leq N-3$, comparing with
the monogamy relation (\ref{CA}), our formula (\ref{MC}) in Theorem 1 gives a tighter
monogamy relation with larger lower bounds. In Theorem 1 we have assumed that
some ${C_{AB_i}}\geq {C_{A|B_{i+1}\cdots B_{N-1}}}$ and some
${C_{AB_j}}\leq {C_{A|B_{j+1}\cdots B_{N-1}}}$ for the $2\otimes2\otimes\cdots\otimes2$ mixed state $\rho\in H_A\otimes H_{B_1}\otimes\cdots\otimes H_{{B_{N-1}}}$.
If all ${C_{AB_i}}\geq {C_{A|B_{i+1}\cdots B_{N-1}}}$ for $i=1, 2, \cdots, N-2$, then we have
the following conclusion:

{\bf [Theorem 2]}.
If ${C_{AB_i}}\geq {C_{A|B_{i+1}\cdots B_{N-1}}}$ for all $i=1, 2, \cdots, N-2$, then we have
\begin{equation}\label{Co}
 C^\alpha_{A|B_1\cdots B_{N-1}}\geq C^\alpha_{A|B_1}+\frac{\alpha}{2} C^\alpha_{A|B_2}+\cdots+\left(\frac{\alpha}{2}\right)^{N-2}C^\alpha_{A|B_{N-1}}.
\end{equation}

{\it Example 1}. Let us consider the three-qubit state $|\psi\rangle$ which can be written in the generalized Schmidt decomposition form \cite{AA,XH},
\be\label{tq}
|\psi\rangle=\lambda_0|000\rangle+\lambda_1e^{i{\varphi}}|100\rangle+\lambda_2|101\rangle
+\lambda_3|110\rangle+\lambda_4|111\rangle,
\ee
where $\lambda_i\geq0,~i=0,\cdots,4$ and $\sum_{i=0}^4\lambda_i^2=1.$
From the definition of concurrence, we have $C_{A|BC}=2\lambda_0\sqrt{{\lambda_2^2+\lambda_3^2+\lambda_4^2}}$, $C_{A|B}=2\lambda_0\lambda_2$, and $C_{A|C}=2\lambda_0\lambda_3$. Set $\lambda_0=\lambda_1=\lambda_2=\lambda_3=\lambda_4=\frac{\sqrt{5}}{5}$. One gets $C_{A|BC}^{\alpha}=(\frac{2\sqrt{3}}{5})^{\alpha}$, $C_{A|B}^{\alpha}+C_{A|C}^{\alpha}=2(\frac{2}{5})^{\alpha}$, $C_{A|B}^{\alpha}+\frac{\alpha}{2}C_{A|C}^{\alpha}
=\left(1+\frac{\alpha}{2}\right)(\frac{2}{5})^{\alpha}$.
The ``residual" entanglement from our result is given by
$y_1=C_{A|BC}^{\alpha}-C_{A|B}^{\alpha}-\frac{\alpha}{2}C_{A|C}^{\alpha}
=(\frac{2\sqrt{3}}{5})^{\alpha}-\left(1+\frac{\alpha}{2}\right)(\frac{2}{5})^{\alpha}$
and the ``residual" entanglement from (\ref{CA}) is
given by $y_2=C_{A|BC}^{\alpha}-C_{A|B}^{\alpha}-C_{A|C}^{\alpha}
=(\frac{2\sqrt{3}}{5})^{\alpha}-2(\frac{2}{5})^{\alpha}$.
One can see that our result is better
than that in \cite{ZXN} for $\alpha\geq2$, see Figure 1.

\begin{figure}
  \centering
  \includegraphics[width=7cm]{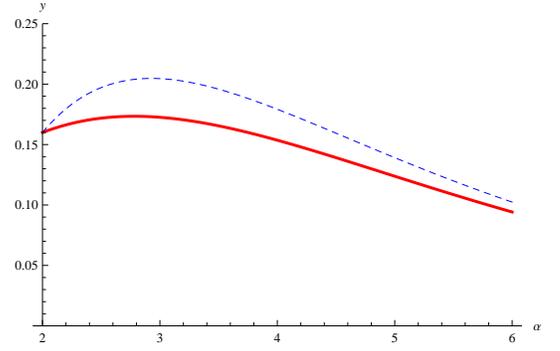}\\
  \caption{$y$ is the ``residual" entanglement as a function of $\alpha$: solid (red) line $y_1$ from our result, dashed (blue) line $y_2$ from the result in \cite{ZXN}.}\label{2}
\end{figure}

We can also derive a tighter upper bound of $C^\alpha_{A|B_1B_2\cdots B_{N-1}}$ for $\alpha<0$.

{\bf [Theorem 3].} For any $2\otimes2\otimes\cdots\otimes2$ mixed state $\rho\in H_A\otimes H_{B_1}\otimes\cdots\otimes H_{B_{N-1}}$ with $C_{AB_i}\neq 0$, $i=1,2,\cdots,N-1$, we have
\begin{equation}\label{TH2}
  C^\alpha_{A|B_1B_2\cdots B_{N-1}}< \tilde{M}(C^\alpha_{A|B_1}+C^\alpha_{A|B_2}+\cdots+C^\alpha_{A|B_{N-1}})
\end{equation}
for all $\alpha<0$, where $\tilde{M}=\frac{1}{N-1}$.

{\sf [Proof].} Similar to the proof of Theorem 1, for arbitrary tripartite state we have
\begin{equation}\label{CON1}
\ba{rcl}
  C^\alpha_{A|B_1B_2}&\leq& \dps (C^2_{AB_1}+C^2_{AB_2})^{\frac{\alpha}{2}}\\
  &=&\dps C^\alpha_{AB_1}\left(1+\frac{C^2_{AB_2}}{C^2_{AB_1}}\right)^{\frac{\alpha}{2}}<C^\alpha_{AB_1},
\ea
\end{equation}
where the first inequality is due to $\alpha<0$ and the second inequality is due to $(1+\frac{C^2_{AB_2}}{C^2_{AB_1}})^{\frac{\alpha}{2}}<1.$
On the other hand, we have
\begin{equation}\label{CON2}
\ba{rcl}
 C^\alpha_{A|B_1B_2}&\leq&\dps (C^2_{AB_1}+C^2_{AB_2})^{\frac{\alpha}{2}}\\
 &=&\dps C^\alpha_{AB_2}\left(1+\frac{C^2_{AB_1}}{C^2_{AB_2}}\right)^{\frac{\alpha}{2}}<C^\alpha_{AB_2}.
\ea
\end{equation}

From (\ref{CON1}) and (\ref{CON2}) we obtain
\begin{equation}\label{XJ}
 C^\alpha_{A|B_1B_2}< \frac{1}{2}(C^\alpha_{AB_1}+C^\alpha_{AB_2}).
\end{equation}
By using the inequality (\ref{XJ}) repeatedly, one gets
\be\label{XCAK}
\ba{l}
 \dps C^{\alpha}_{A|B_1B_2\cdots B_{N-1}}< \frac{1}{2}(C^{\alpha}_{A|B_1}+C^{\alpha}_{A|B_2\cdots B_{N-1}})
\\[3mm]
 \dps~~~~ <\frac{1}{2}C^{\alpha}_{A|B_1}+\left(\frac{1}{2}\right)^2C^{\alpha}_{A|B_2}
 +\left(\frac{1}{2}\right)^2C^{\alpha}_{A|B_3\cdots B_{N-1}}\\[3mm]
 \dps~~~~<\cdots< \frac{1}{2}C^{\alpha}_{A|B_1}+\left(\frac{1}{2}\right)^2C^{\alpha}_{A|B_2}
  +\cdots\\[3mm]
 \dps~~~~~~~~~~~~~~~~+\left(\frac{1}{2}\right)^{N-2}C^{\alpha}_{A|B_{N-2}} +\left(\frac{1}{2}\right)^{N-2} C^{\alpha}_{A|B_{N-1}}.
\ea
\ee
By cyclically permuting the sub-indices $B_1, B_2, \cdots, B_{N-1}$ in (\ref{XCAK}) we can get
a set of inequalities. Summing up these inequalities we have (\ref{TH2}).
\qed

As the factor $\tilde{M}=\frac{1}{N-1}$ is less than one,
the inequality (\ref{TH2}) is tighter than the one in \cite{ZXN}.
This factor $\tilde{M}$ depends on the number of partite $N$.
Namely, for larger multipartite systems, the inequality (\ref{TH2}) gets even tighter
than the one in \cite{ZXN}.

{\it Example 2}. Let us consider again the three-qubit state (\ref{tq}). In this case, we have $N=3$ and $\tilde{M}={1}/{2}$. Taking the same parameters used in Example 1, we have
$C_{A|BC}^{\alpha}=(\frac{2\sqrt{3}}{5})^{\alpha}$, $C_{A|B}^{\alpha}+C_{A|C}^{\alpha}=2(\frac{2}{5})^{\alpha}$, $\tilde{M}(C_{A|B}^{\alpha}+C_{A|C}^{\alpha})=(\frac{2}{5})^{\alpha}.$
Comparing the function of $y_1=C_{A|BC}^{\alpha}-\tilde{M}C_{A|B}^{\alpha}-\tilde{M}C_{A|C}^{\alpha}
=(\frac{2\sqrt{3}}{5})^{\alpha}-(\frac{2}{5})^{\alpha}$ with $y_2=C_{A|BC}^{\alpha}-C_{A|B}^{\alpha}-C_{A|C}^{\alpha}
=(\frac{2\sqrt{3}}{5})^{\alpha}-2(\frac{2}{5})^{\alpha}$, one can see that
our result is better than the one from \cite{ZXN}, see Figure 2.

\begin{figure}
  \centering
  \includegraphics[width=7cm]{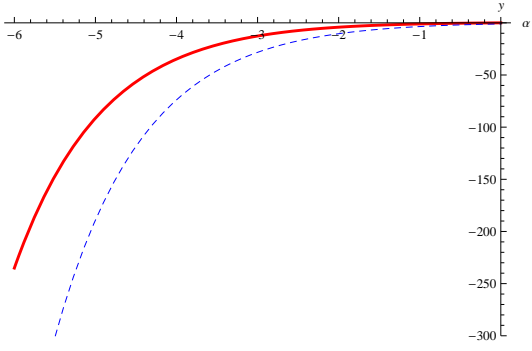}\\
  \caption{$y$ is the ``residual" entanglement as a function of $\alpha$:
   red line (solid line) from our Theorem 2; blue line (dashed line ) from the result in \cite{ZXN}.}\label{3}
\end{figure}

{\it [Remark]}~ In (\ref{TH2}) we have assumed that all $C_{AB_i}$, $i=1,2,\cdots,N-1$, are nonzero.
In fact, if one of them is zero, the inequality still holds if one removes this term from the inequality. Namely, if $C_{AB_i}=0,$ then one has $C^{\alpha}_{A|B_1B_2\cdots B_{N-1}}<\frac{1}{2}C^{\alpha}_{A|B_1}+\cdots+\left(\frac{1}{2}\right)^{i-1}C^{\alpha}_{A|B_{i-1}}+\left(\frac{1}{2}\right)^{i}C^{\alpha}_{A|B_{i+1}}+\cdots+\left(\frac{1}{2}\right)^{N-3}C^{\alpha}_{A|B_{N-2}}+\left(\frac{1}{2}\right)^{N-3} C^{\alpha}_{A|B_{N-1}}$. Similar to the analysis in proving Theorem 2, one gets $C^\alpha_{A|B_1B_2\cdots B_{N-1}}<\frac{1}{N-1}
(C^\alpha_{A|B_1}+\cdots+C^\alpha_{A|B_{i-1}}+C^\alpha_{A|B_{i+1}}+\cdots+C^\alpha_{A|B_{N-1}}),
$ for $\alpha<0$.

\section{TIGHTER MONOGAMY INEQUALITY FOR EoF}

The entanglement of formation (EoF) \cite{CHHJ,CHDP} is a well defined important measure of entanglement for bipartite systems. Let $H_A$ and $H_B$ be $m$ and $n$ dimensional $(m\leq n)$ vector spaces, respectively. The EoF of a pure state $|\psi\rangle\in H_A\otimes H_B$ is defined by
\begin{equation}\label{SEA}
 E(|\psi\rangle)=S(\rho_A),
\end{equation}
where $\rho_A=\mathrm{Tr}_B(|\psi\rangle\langle\psi|)$ and $S(\rho)=-\mathrm{Tr}(\rho \log_2\rho).$ For a bipartite mixed state $\rho_{AB}\in H_A\otimes H_B$, the entanglement of formation is given by
\begin{equation}\label{SEB}
 E(\rho_{AB})=\min_{\{p_i,|\psi_i\rangle\}}\sum_ip_iE(|\psi_i\rangle)
\end{equation}
with the minimum taking over all possible decompositions of $\rho_{AB}$ in a mixture of pure states $\rho_{AB}=\sum_ip_i|\psi_i\rangle\langle\psi_i|$, where $p_i\geq0$ and $\sum_ip_i=1$.

Denote $f(x)=H\left(\frac{1+\sqrt{1-x}}{2}\right)$, where $H(x)=-x\log_2(x)-(1-x)\log_2(1-x)$. From (\ref{SEA}) and (\ref{SEB}), one has $E(|\psi\rangle)=f\left(C^2(|\psi\rangle)\right)$ for $2\otimes m~(m\geq2)$ pure state $|\psi\rangle$, and  $E(\rho)=f\left(C^2(\rho)\right)$ for two-qubit mixed state $\rho$ \cite{WK}. It is obvious that $f(x)$ is a monotonically increasing function for $0\leq x\leq1$. $f(x)$ satisfies the following relations:
\begin{equation}\label{F2}
  f^{\sqrt{2}}(x^2+y^2)\geq f^{\sqrt{2}}(x^2)+f^{\sqrt{2}}(y^2),
\end{equation}
where $f^{\sqrt{2}}(x^2+y^2)=[f(x^2+y^2)]^{\sqrt{2}}.$

It has been show that the entanglement of formation does not satisfy the inequality $E_{AB}+E_{AC}\leq E_{A|BC}$ \cite{VC}. In \cite{YKN} the authors showed that EoF is a monotonic function $E^2(C^2_{A|B_1B_2\cdots B_{N-1}})\geq E^2(\sum_{i=1}^{N-1}C^2_{AB_i})$. It is further proved that for $N-$qubit systems, one has \cite{ZXN}
\begin{equation}\label{zxneof}
 E^\alpha_{A|B_1B_2\cdots B_{N-1}}\geq E^\alpha_{A|B_1}+E^\alpha_{A|B_2}+\cdots+E^\alpha_{A|B_{N-1}}
\end{equation}
for $\alpha\geq\sqrt{2}$, where $E_{A|B_1B_2\cdots B_{N-1}}$ is the entanglement of formation of $\rho$ in bipartite partition $A|B_1B_2\cdots B_{N-1}$, and $E_{AB_i}$, $i=1,2,\cdots,N-1$, is the entanglement of formation of the mixed states $\rho_{AB_i}=\mathrm{Tr}_{B_1B_2\cdots B_{i-1},B_{i+1}\cdots B_{N-1}}(\rho)$.
In fact, generally we can prove the following results.

{\bf [Theorem 4].} For any N-qubit mixed state $\rho\in H_A\otimes H_{B_1}\otimes\cdots\otimes H_{{B_{N-1}}}$, if
${C_{AB_i}}\geq {C_{A|B_{i+1}\cdots B_{N-1}}}$ for $i=1, 2, \cdots, m$, and
${C_{AB_j}}\leq {C_{A|B_{j+1}\cdots B_{N-1}}}$ for $j=m+1,\cdots,N-2$, $\forall$ $1\leq m\leq N-3$, $N\geq 4$, the entanglement of formation $E(\rho)$ satisfies
\begin{eqnarray}\label{EA}
  E^\alpha_{A|B_1B_2\cdots B_{N-1}}&\geq& E^\alpha_{A|B_1}+t E^\alpha_{A|B_2}\cdots+t^{m-1}E^\alpha_{A|B_m}\nonumber\\
 &&+t^{m+1}(E^\alpha_{A|B_{m+1}}+\cdots+E^\alpha_{A|B_{N-2}})\nonumber\\
 &&+t^{m}E^\alpha_{A|B_{N-1}},
\end{eqnarray}
for $\alpha\geq\sqrt{2}$, where $t={\alpha}/{\sqrt{2}}$.

{\sf [Proof].} For $\alpha\geq\sqrt{2}$, we have
\begin{eqnarray}\label{FA}
 f^{{\alpha}}(x^2+y^2)&&=\left(f^{\sqrt{2}}(x^2+y^2)\right)^t \nonumber\\
 &&\geq \left(f^{\sqrt{2}}(x^2)+f^{\sqrt{2}}(y^2)\right)^t \\
 &&\geq \left(f^{\sqrt{2}}(x^2)\right)^t+t\left(f^{\sqrt{2}}(y^2)\right)^t\nonumber\\
 &&=f^{\alpha}(x^2)+tf^{\alpha}(y^2),\nonumber
\end{eqnarray}
where the first inequality is due to the inequality (\ref{F2}), and the second inequality is obtained from a similar consideration in the proof of the second inequality in (\ref{CK1}).

Let $\rho=\sum_ip_i|\psi_i\rangle\langle\psi_i|\in H_A\otimes H_{B_1}\otimes\cdots\otimes H_{{B_N-1}}$ be the optimal decomposition of $E_{A|B_1B_2\cdots B_{N-1}}(\rho)$ for the N-qubit mixed state $\rho$, we have
\begin{eqnarray*}\label{ED}
&&E_{A|B_1B_2\cdots B_{N-1}}(\rho)\nonumber\\
&&=\sum_ip_iE_{A|B_1B_2\cdots B_{N-1}}(|\psi_i\rangle)\nonumber\\
&&=\sum_ip_if\left(C^2_{A|B_1B_2\cdots B_{N-1}}(|\psi_i\rangle)\right)\nonumber\\
&&\geq f\left(\sum_ip_iC^2_{A|B_1B_2\cdots B_{N-1}}(|\psi_i\rangle)\right)\nonumber\\
&&\geq f\left(\left[\sum_ip_iC_{A|B_1B_2\cdots B_{N-1}}(|\psi_i\rangle)\right]^2\right)\nonumber\\
&& \geq f\left(C^2_{A|B_1B_2\cdots B_{N-1}}(\rho)\right),\nonumber
\end{eqnarray*}
where the first inequality is due to that $f(x)$ is a convex function. The second inequality is due to the Cauchy-Schwarz inequality: $(\sum_ix_i^2)^{\frac{1}{2}}(\sum_iy_i^2)^{\frac{1}{2}}\geq\sum_ix_iy_i$, with $x_i=\sqrt{p_i}$ and $y_i=\sqrt{p_i}C_{A|B_1B_2\cdots B_{N-1}}(|\psi_i\rangle)$. Due to the definition of concurrence and that $f(x)$ is a monotonically increasing function, we obtain the third inequality.
Therefore, we have
\begin{eqnarray*}
&&E^\alpha_{A|B_1B_2\cdots B_{N-1}}(\rho)\nonumber\\
&&\geq f^\alpha(C^2_{AB_1}+C^2_{AB_2}+\cdots+C^2_{AB_{m-1}})\nonumber\\
&&\geq f^{\alpha}(C^2_{A|B_1})+t f^{\alpha}(C^2_{A|B_2})\cdots+t^{m-1} f^{\alpha}(C^2_{A|B_m})\nonumber\\
&&~~~+t^{m+1}\left(f^{\alpha}(C^2_{A|B_{m+1}})+\cdots+f^{\alpha}(C^2_{A|B_{N-2}})\right)\nonumber\\
&&~~~+t^{m}f^{\alpha}(C^2_{A|B_{N-1}})\nonumber\\
&&=E^\alpha_{A|B_1}+t E^\alpha_{A|B_2}\cdots+t^{m-1}E^\alpha_{A|B_m}\nonumber\\
 &&~~~+t^{m+1}(E^\alpha_{A|B_{m+1}}+\cdots+E^\alpha_{A|B_{N-2}})
 +t^{m}E^\alpha_{A|B_{N-1}},\nonumber
\end{eqnarray*}
where we have used the monogamy inequality in (\ref{C2}) for $N-$qubit states $\rho$ to obtain the first
inequality. By using (\ref{FA}) and the similar consideration in the proof of Theorem 1, we get the second
inequality. Since for any $2\otimes2$ quantum state $\rho_{AB_i}$, $E(\rho_{AB_i})=f\left[C^2(\rho_{AB_i})\right]$, one gets the last equality. \qed

As the factor $t={\alpha}/{\sqrt{2}}$ is greater or equal to one for $\alpha\geq\sqrt{2}$,
(\ref{EA}) is obviously tighter than (\ref{zxneof}). Moreover,
similar to the concurrence, for the case that ${C_{AB_i}}\geq {C_{A|B_{i+1}\cdots B_{N-1}}}$ for all $i=1, 2, \cdots, N-2$, we have a simple tighter monogamy relation for entanglement of formation:

{\bf [Theorem 5]}.
If ${C_{AB_i}}\geq {C_{A|B_{i+1}\cdots B_{N-1}}}$ for all $i=1, 2, \cdots, N-2$, we have
\begin{equation}\label{coeof}
\ba{rcl}
 E^\alpha_{A|B_1B_2\cdots B_{N-1}}&\geq& \dps E^\alpha_{A|B_1}+\frac{\alpha}{\sqrt{2}} E^\alpha_{A|B_2}+\cdots\\[4mm]
&&\dps +\left(\frac{\alpha}{\sqrt{2}}\right)^{N-2}E^\alpha_{A|B_{N-1}}
\ea
\end{equation}
for $\alpha\geq\sqrt{2}$.

{\it Example 3}. Let us consider the $W$ state, $|W\rangle=\frac{1}{\sqrt{3}}(|100\rangle+|010\rangle+|001\rangle).$ We have $E_{AB}=E_{AC}=0.55$, $E_{A|BC}=0.92$. Let $y_1=E_{A|BC}^{\alpha}-E_{A|B}^{\alpha}-\frac{\alpha}{\sqrt{2}}E_{A|C}^{\alpha}$
denote the residual entanglement from our formula (\ref{coeof}),
and $y_2=E_{A|BC}^{\alpha}-E_{A|B}^{\alpha}-E_{A|C}^{\alpha}$
the residual entanglement from formula (\ref{zxneof}). It is easily verified that
our results is better than the one in \cite{ZXN} for $\alpha\geq\sqrt{2}$, see Figure 3.

\begin{figure}
\centering
    \includegraphics[width=7cm]{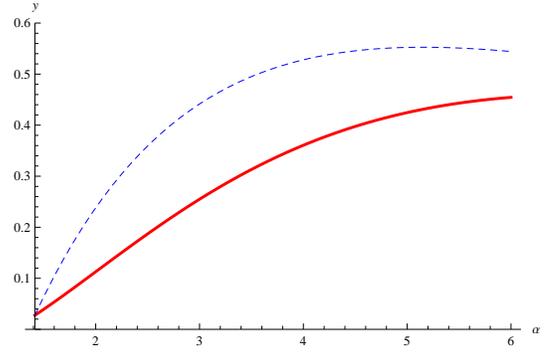}\\
  \caption{$y$ is the residual entanglement as a function of $\alpha$:
   red (solid) line from our results; blue (dashed) line from the result in \cite{ZXN}.}\label{3}
\end{figure}

\section{CONCLUSION}

Entanglement monogamy is a fundamental property of multipartite entangled states. We have investigated the monogamy relations related to the concurrence and EoF, and presented tighter entanglement monogamy relations of $C^{\alpha}$ and $E^{\alpha}$ for $\alpha\geq2$ and $\alpha\geq\sqrt{2}$, respectively.
Monogamy relations characterize the distributions of entanglement in multipartite systems.
Tighter monogamy relations imply finer characterizations of the entanglement distribution.
Our approach may be also used to study further the monogamy properties related to other quantum entanglement measures such as negativity and quantum correlations such as quantum discord.

\bigskip
\noindent{\bf Acknowledgments}\, \,
We thank Bao-Zhi Sun, Xue-Na Zhu and Xian Shi for very useful discussions.
This work is supported by NSFC under number 11275131.


\bigskip

\begin{thebibliography}{99}
\bibitem{MANC} M. A. Nielsen and I. L. Chuang, Quantum Computation and Quantum Information, Cambridge: Cambridge University Press, 2000.
\bibitem{HPMK} R. Horodecki, P. Horodecki, M. Horodecki, and K. Horodecki, Quantum entanglement.
Rev. Mod. Phys. 81, 865 (2009).
\bibitem{FMA} F. Mintert, M. Ku\'s, and A. Buchleitner, Concurrence of Mixed Bipartite Quantum States in Arbitrary Dimensions. Phys. Rev. Lett. 92, 167902 (2004).
\bibitem{CAF} K. Chen, S. Albeverio, and S. M. Fei, Concurrence of Arbitrary Dimensional Bipartite Quantum States. Phys. Rev. Lett. 95, 040504 (2005).
\bibitem{HPBS}H. P. Breuer, Separability criteria and bounds for entanglement measures. J. Phys. A: Math. Gen. 39, 11847 (2006).
\bibitem{HPBO} H. P. Breuer, Optimal Entanglement Criterion for Mixed Quantum States. Phys. Rev. Lett. 97, 080501 (2006).
\bibitem{JIDV} J. I. de Vicente, Lower bounds on concurrence and separability conditions. Phys. Rev. A 75, 052320 (2007).
\bibitem{ZZZG} C. J. Zhang, Y. S. Zhang, S. Zhang, and G. C. Guo, Optimal entanglement witnesses based on local orthogonal observables. Phys. Rev. A 76, 012334 (2007).
\bibitem{MP} M. Pawlowski, Security proof for cryptographic protocols based only on the monogamy of Bell’s inequality violations. Phys. Rev. A 82, 032313 (2010).	
\bibitem{MK} M.  Koashi  and  A.  Winter, Monogamy of quantum entanglement and other correlations. Phys.  Rev.  A  69,  022309 (2004).
\bibitem{TJ} T. J. Osborne and F. Verstraete, General Monogamy Inequality for Bipartite Qubit Entanglement. Phys. Rev. Lett. 96, 220503 (2006).	
\bibitem{YKM} Y. K. Bai, M. Y. Ye, and Z. D. Wang, Entanglement monogamy and entanglement evolution in multipartite systems. Phys. Rev. A 80, 044301 (2009).	
\bibitem{TR} T. R. de Oliveira, M. F. Cornelio, and F. F. Fanchini, Monogamy of entanglement of formation. Phys. Rev. A 89, 034303 (2014).	
\bibitem{AU} A. Uhlmann, Fidelity and concurrence of conjugated states. Phys. Rev. A 62, 032307 (2000).	
\bibitem{ZXN} X. N. Zhu and S. M. Fei, Entanglement monogamy relations of qubit systems. Phys. Rev. A 90, 024304 (2014).
\bibitem{PR} P. Rungta, V. Buzek, C. M. Caves, M. Hillery, and G. J. Milburn, Universal state inversion and concurrence in arbitrary dimensions. Phys. Rev. A 64, 042315 (2001).
\bibitem{SA} S. Albeverio and S. M. Fei, A note on invariants and entanglements. J. Opt. B: Quantum Semiclass Opt. 3, 223 (2001).	
\bibitem{XJ} X. J. Ren and W. Jiang, Entanglement monogamy inequality in a $2\otimes 2\otimes4$ system. Phys. Rev. A 81, 024305 (2010).

\bibitem{AA} A. Acin, A. Andrianov, L. Costa, E. Jane, J. I. Latorre, and R. Tarrach, Generalized Schmidt Decomposition and Classification of Three-Quantum-Bit States. Phys. Rev. Lett. 85, 1560 (2000).
\bibitem{XH}  X. H. Gao and S. M. Fei, Estimation of concurrence for multipartite mixed states. Eur. Phys. J. Special Topics 159, 71-77 (2008).


\bibitem{CHHJ} C. H. Bennett, H. J. Bernstein, S. Popescu, and B. Schumacher, Concentrating partial entanglement by local operations. Phys. Rev. A 53, 2046 (1996).
\bibitem{CHDP} C. H. Bennett, D. P. DiVincenzo, J. A. Smolin, and W. K. Wootters, Mixed-state entanglement and quantum error correction. Phys. Rev. A 54, 3824 (1996).
\bibitem{WK} W. K. Wootters, Entanglement of Formation of an Arbitrary State of Two Qubits. Phys. Rev. Lett. 80, 2245 (1998).
\bibitem{VC} V. Coffman, J. Kundu, and W. K. Wootters, Distributed entanglement. Phys. Rev. A 61, 052306 (2000).
\bibitem{YKN} Y. K. Bai, N. Zhang, M. Y. Ye, and Z. D. Wang, Exploring multipartite quantum correlations with the square of quantum discord. Phys. Rev. A 88, 012123 (2013).


\end{thebibliography}
\end{document}